\documentclass[twocolumn,9pt]{article}
\usepackage{extsizes}
\usepackage[super,sort&compress,comma]{natbib} 
\usepackage[version=3]{mhchem}
\usepackage[left=1.5cm, right=1.5cm, top=2.0cm, bottom=2.0cm]{geometry}
\usepackage{balance}
\usepackage{mathptmx}
\usepackage{sectsty}
\usepackage{graphicx} 
\usepackage{lastpage}
\usepackage[format=plain,justification=justified,singlelinecheck=false,font={stretch=1.125,small,sf},labelfont=bf,labelsep=space]{caption}
\usepackage{float}
\usepackage[english]{babel}
\addto{\captionsenglish}{%
  
}
\usepackage{array}
\usepackage{droidsans}
\usepackage{charter}
\usepackage[T1]{fontenc}
\usepackage[usenames,dvipsnames]{xcolor}
\usepackage{setspace}
\usepackage[compact]{titlesec}
\usepackage{hyperref}
\usepackage{enumitem}
\usepackage{booktabs}
\usepackage{epstopdf}

\usepackage{amssymb}

\begin{document}

\makeatletter
\renewcommand\LARGE{\@setfontsize\LARGE{15pt}{17}}
\renewcommand\Large{\@setfontsize\Large{12pt}{14}}
\renewcommand\large{\@setfontsize\large{10pt}{12}}
\renewcommand\footnotesize{\@setfontsize\footnotesize{7pt}{10}}
\makeatother
\renewcommand{\thefootnote}{\fnsymbol{footnote}}
\setcounter{secnumdepth}{5}
\makeatletter 
\renewcommand\@biblabel[1]{#1}            
\renewcommand\@makefntext[1]%
{\noindent\makebox[0pt][r]{\@thefnmark\,}#1}
\makeatother 
\renewcommand{\figurename}{\small{Fig.}~}
\sectionfont{\sffamily\Large}
\subsectionfont{\normalsize}
\subsubsectionfont{\bf}
\setstretch{1.125}
\setlength{\skip\footins}{0.8cm}
\setlength{\footnotesep}{0.25cm}
\setlength{\jot}{10pt}
\titlespacing*{\section}{0pt}{4pt}{4pt}
\titlespacing*{\subsection}{0pt}{15pt}{1pt}
\setlength{\arrayrulewidth}{1pt}
\setlength{\columnsep}{6.5mm}
\setlength\bibsep{1pt}

\twocolumn[
  \begin{@twocolumnfalse}
  \vspace{1em}
  \sffamily
  \noindent\LARGE{\textbf{A General Pipeline for Digesting Scientific Literature into a Shared Scientific Knowledge Base}}
  \vspace{0.5em}
  
  \noindent\large{C.T.\ Black$^{\ast}$\textit{$^{a}$}}
  \vspace{0.5em}

  \noindent\normalsize{The published scientific literature is a rich, continuously growing record of measurements, correlations, and observations that modern AI tools can now make accessible in new ways. The Materials Explorer Pipeline digests collections of scientific papers into a structured, queryable database, producing sample records with full provenance and confidence, making them interactively explorable, and surfacing hypothesis candidates for scientist review. Each extracted record is a self-contained, portable unit of knowledge, carrying the measurements, research details, and source citations needed to use and cite the data appropriately. The Pipeline is demonstrated on recent superconducting qubit materials literature of the Co-design Center for Quantum Advantage, a DOE National Quantum Information Science Research Center, producing a corpus of 233 samples across 10 material classes. The Pipeline architecture is domain-agnostic and designed to be readily portable to other scientific domains.}
  \vspace{0.6cm}
  \end{@twocolumnfalse}
]

\renewcommand*\rmdefault{bch}\normalfont\upshape
\rmfamily
\section*{}
\vspace{-1cm}

\footnotetext{\textit{$^{a}$~Co-design Center for Quantum Advantage, Brookhaven National Laboratory, Upton, New York 11973, USA. E-mail: ctblack@bnl.gov}}
\footnotetext{\dag~Supplementary Information available. See DOI: 00.0000/00000000.}

\section*{Introduction}
The published scientific literature has always been the primary record of what a field knows. The measurements and observations reported in peer-reviewed publications are a rich scientific resource that is already validated and citable. Word embedding studies have shown that latent useful scientific knowledge is encoded in this published record, pointing toward the possibility of extracting it systematically and at scale. \cite{tshitoyan2019} For scientists, working with the vast scientific literature in aggregate is a growing challenge. Results reported across hundreds of papers from dozens of groups, using varied terminology and formats, are difficult to compare, and no shared resource exists to support cross-group comparisons or systematic hypothesis testing against the full published record. 

Here, we describe the Materials Explorer Pipeline, which gives scientists a new way to visualize and make use of their collective knowledge, presenting measurements and derived quantities for further use, and surfacing correlations and hypothesis candidates to evaluate.  In this way, the Pipeline is part of a broader opportunity for AI tools to extend scientific cognition. \cite{yager2024}

The Pipeline sidesteps an impediment to community database adoption, which requires scientists to contribute data separately from publication, a task that most do not take on. Computationally-derived databases such as the Materials Project\cite{jain2013} have shown the high value of structured, queryable materials data at scale, but experimental property databases of comparable scope remain scarce. The Materials Explorer Pipeline can draw from papers already produced. A schema encoding the measurements of interest, paired with an automated ingestion pipeline, populates a structured database from the existing body of literature, with no burden on the contributors. Human review is built into every stage where AI makes an inference, ensuring that the digested corpus reflects scientific judgment. Throughout, data extraction is performed via repeated calls to a large language model.

Early approaches to automated extraction of structured data from scientific publications framed the problem as named entity recognition on sentence fragments, achieving strong benchmark performance on materials science text but not addressing  assembly of complete, multi-field records from full-length documents.\cite{weston2019} A significant advance later came from showing that fine-tuned language models can extract schema-defined records from materials science texts,\cite{dagdelen2024} with the limitation that fine-tuning required hundreds of manually annotated examples per task and locked the schema at training time. Subsequent work showed that zero-shot prompting of frontier models can achieve comparable extraction quality without labeled training data.\cite{polak2024,ansari2023} A comprehensive review of this approach for chemistry and materials science, covering prompting strategies, multimodal methods, and evaluation, is available in the literature.\cite{schillingwilhelmi2025}  Most recently, a tiered extraction hierarchy using lightweight relevance triage before full structured extraction was applied to experimental records in shock physics, \cite{rameshbabu2026} an approach similar to the one employed in this work.  The Materials Explorer Pipeline builds on this research progression, introducing a mandatory catchall block to the schema that captures knowledge not yet formalized into named fields, which allows the schema to grow more capable as the corpus accumulates.

\begin{figure}[h]
\centering
\includegraphics[width=\columnwidth]{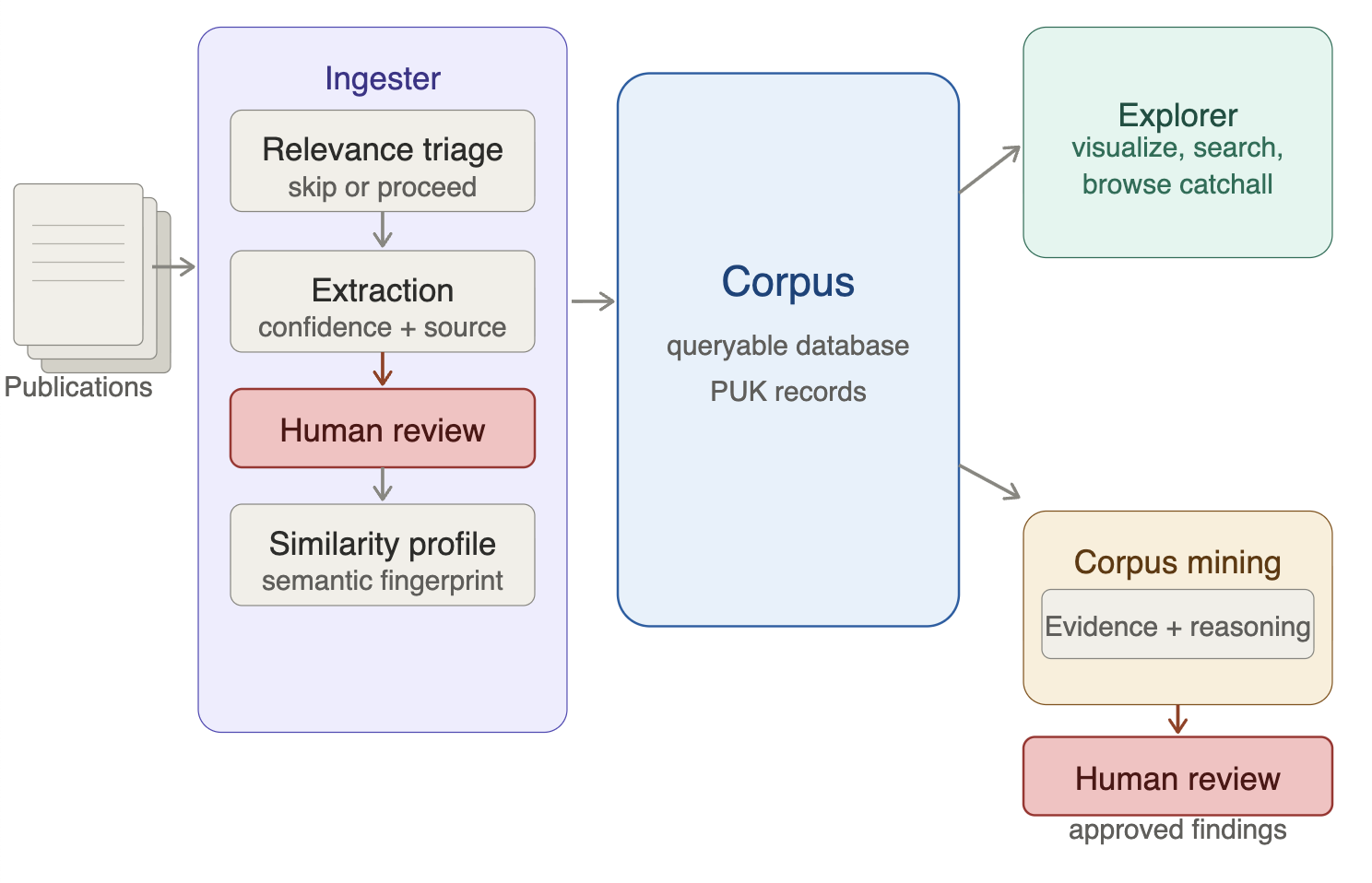}
\caption{\textbf{Materials Explorer Pipeline architecture}. Publications are digested by the Ingester, which analyzes each paper and populates the Corpus. The Explorer makes the Corpus interactively accessible for visualization, search, and browsing. Corpus Mining surfaces candidate scientific hypotheses with supporting and complicating evidence. Human reviews and approvals appear after ingestion and AI reasoning in corpus mining.}
\label{fig:pipeline}
\end{figure}

The Pipeline has four components that work together to build a shared corpus of scientific knowledge and make it useful (Fig.~\ref{fig:pipeline}).  The database schema guides knowledge encoding from the publications, determining the most important information to capture and enabling the downstream comparisons that give the data value. The Ingester analyzes each paper and populates the database. The browser-based Explorer visualizes the corpus and renders it interactively accessible. The corpus mining tool operates over the complete set of records, surfacing candidate scientific hypotheses together with supporting and complicating evidence. The corpus database is the central artifact the Pipeline produces, fully traceable to the source publications and growing continually as new papers are ingested.  This Pipeline is demonstrated using superconducting qubit materials literature published in the last year by the Co-design Center for Quantum Advantage (C2QA) \cite{c2qa}, a DOE National Quantum Information Science Research Center, though the Pipeline architecture is domain-agnostic, as described below.
\section*{Schema Design}
The database schema is the intellectual heart of the Pipeline, encoding domain knowledge such that the sample information captured from each publication becomes immediately useful. The schema defines the initial determinations of what information gets recorded and thereby which comparisons across publications are enabled.

\begin{figure}[h]
\centering
\includegraphics[width=\columnwidth]{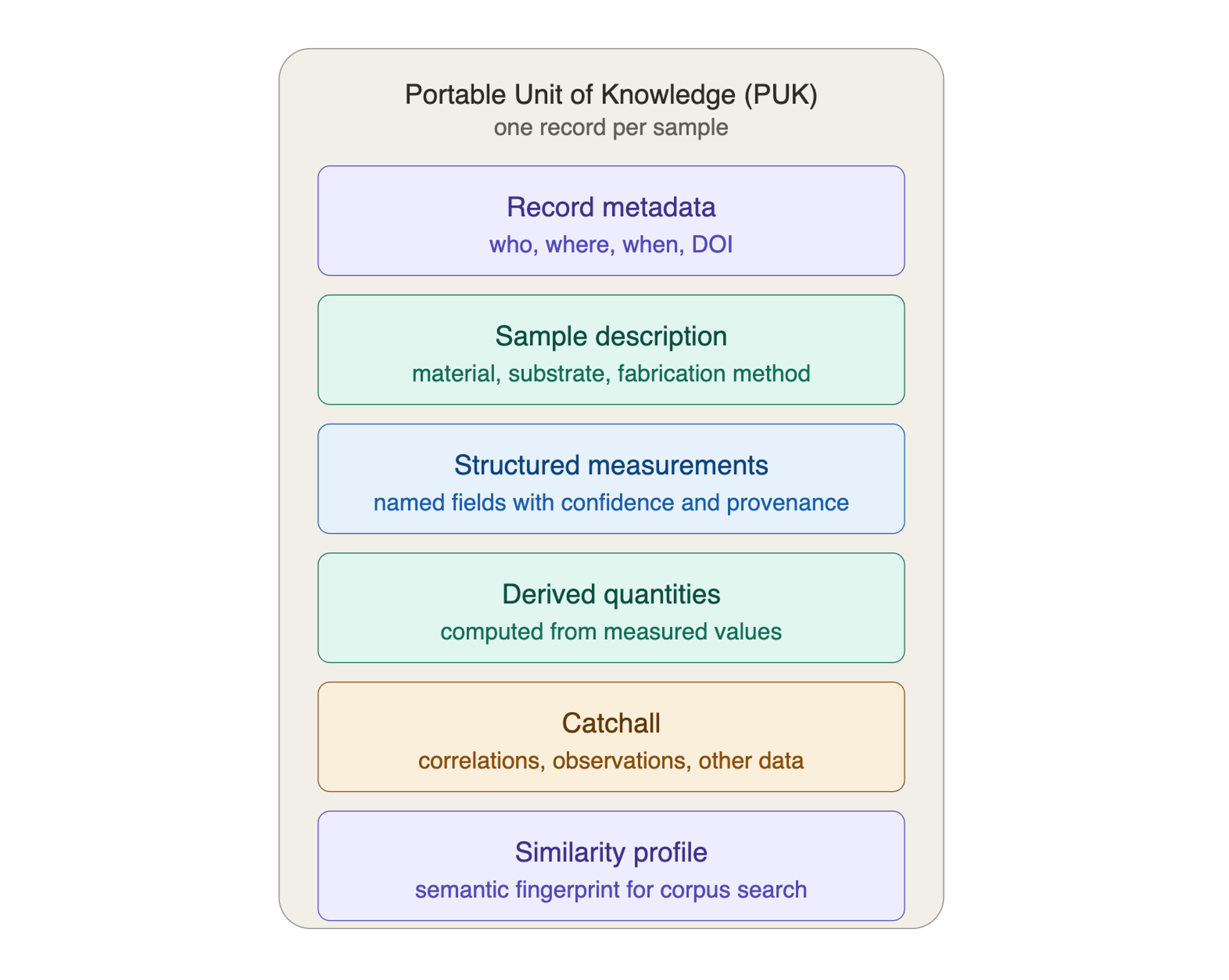}
\caption{The six-block schema of a Portable Unit of Knowledge (PUK). The record metadata establishes provenance. The sample description provides material and processing information. Structured measurements are reported values with confidence levels. Derived quantities are computed from measured values using standard formulas. The catchall holds information not covered by named schema fields. The similarity profile is a semantic sample fingerprint.}
\label{fig:schema}
\end{figure}

Each database record corresponds to a single sample characterized in a publication and has a six-block structure: metadata, sample description, measurements, derived quantities, catchall, and similarity profile. Single papers can thus contribute many sample records. Together, these six blocks make each record a self-contained Portable Unit of Knowledge (PUK) (Fig.~\ref{fig:schema}), which carries everything needed to understand, use, and appropriately cite a sample's data independently of the Pipeline that produced it, and enabling records to move freely between institutions and research groups. A complete example PUK is in the Supplementary Information (S1).  The metadata and sample description establish the identity and provenance of the record, including what material was studied, how it was made, and where the data came from. The measurements block records reported values, with derived quantities calculated from reported values using standard formulas. The Block~3 field definitions are provided in the Supplementary Information (S2). A catchall block holds relevant information that does not fit any current schema field. Each record also contains a similarity profile that enables corpus-wide search and comparison.

Prior systems typically discard information that does not fit the schema at design time. The Pipeline's catchall block is an architecture for capturing author-stated correlations, anomalous observations, and additional measurements. Fields appearing in the catchall across more than 5\% of corpus samples are presented for review as potential candidates for promotion to named schema columns, after human approval. In this way, the schema evolves to better represent what the corpus contains as it grows.
\section*{The Ingester}
The publications Ingester processes papers as PDF files, submitting them to the large language model (LLM) for structured data extraction from figures, tables, and manuscript text. This makes use of the model's built-in multimodal document understanding, obviating the need for a PDF parser, OCR pipeline, or figure extraction code.  Publications were sourced from open repositories arXiv and DOE PAGES.\cite{arxiv,doepages} The Pipeline produces sparse output containing only schema fields reported in the paper, with no paper text or figures reproduced in the corpus.  The Ingester proceeds in three stages: triage for topical relevance, structured paper digestion, and generation of sample similarity profiles.

In the first stage, an LLM evaluates each paper for relevance, using a prompt that specifies the materials and measurement types of interest. Those determined to be not relevant are logged in an idempotency ledger and skipped. This lightweight triage takes ~15 seconds per paper and economizes on processing. 

Papers that pass triage proceed to structured digestion, in which the LLM processes the full document using a prompt structured around the schema fields, creating and populating entries for each sample found in the paper. The prompt also directs the LLM to populate each record's catchall block with information that does not fit any named schema field. Full digestion can take up to several minutes, depending on the number of samples found. 

In the third stage, the LLM generates a similarity profile for each sample, characterizing it across a set of pre-defined scientific categories (See example profile in Supplementary Information (S1)). These profiles enable corpus-wide search and comparisons of scientifically similar samples.

Two additional design decisions support reliable operation at scale. The Ingester maintains a processed ledger keyed by a stable publication identifier (Digital Object Identifier (DOI) or arXiv number) so previously processed publications are recognized even if renamed or moved. The corpus is stored as an append-only ledger (JSONL), from which the Pipeline derives a database for queries and visualizations. Suspected duplicates are identified and presented for a ``keep-or-discard'' decision, lessening the chance that the corpus becomes polluted with duplicate records of the same work.
\section*{The Explorer}
\begin{figure*}[h]
\centering
\includegraphics[width=0.7\textwidth]{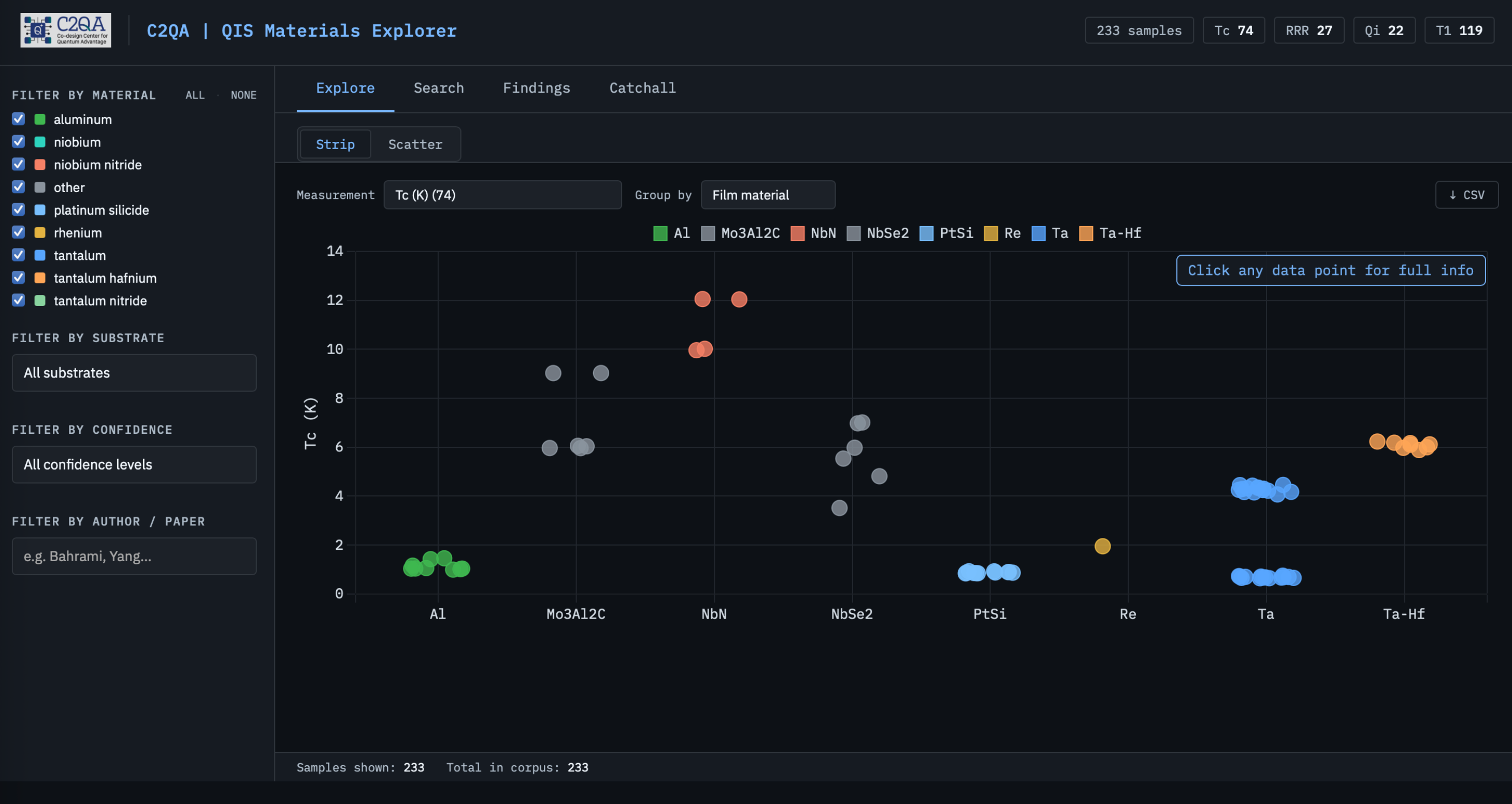}
\caption{Explorer screenshot showing the Explore tab with strip plot view of the C2QA corpus, grouped by material.}
\label{fig:explorer}
\end{figure*}

The Explorer user interface provides browser-based access to the full sample corpus, for scientists to visually interact with the dataset, perform searches, and make comparisons between samples. The Explorer plots any user-selected schema field across the full record set, filters by material class, and compares between material systems in a single view (Fig.~\ref{fig:explorer}). When a value is reported in multiple ways, the Explorer displays a single best-available value, keeping visualizations scientifically meaningful without requiring the user to adjudicate between variants. Each data point on any plot is clickable, and opens to reveal the full sample record, giving ready, downloadable access to all measurements, derived values, processing details, and source citations.

The Explorer supports similarity search for records most similar to a sample of interest, as evaluated by a hybrid metric that combines similarity of measured values and semantic profiles. This enables discovery of related work across the corpus. For example, samples from different research groups, fabrication approaches, or time periods may share scientific character, but may not surface when filtering on any single measurement value.

The Explorer also provides browseable and searchable access to the unformalized knowledge stored in the corpus catchall fields, including author-stated correlations and anomalous observations not captured in named schema fields, and displays human-approved corpus mining findings with supporting and complicating evidence.
\section*{Corpus Mining and Scientific Hypothesis Generation}
In addition to measurements and data, the literature of any active research field contains years of accumulated scientific judgment, including observations, correlations, and hypotheses reported across hundreds of papers from dozens of research groups. Modern tools make it possible to work with this accumulated knowledge in new ways, assembling evidence across the full weight of the published record in ways that were not previously practical. The Materials Explorer Pipeline surfaces hypothesis candidates by drawing from what scientists have already reported in publications, tests them against available evidence, and provides summary findings for human evaluation.

This process begins with a mechanical read of all correlation entries from the corpus catchall blocks, which maps descriptive terms to canonical schema field names and builds co-occurrence evidence tables showing how often paired measurements appear together across the corpus.
The Pipeline creates two types of evidence tables.  One reads across the entire corpus with all material classes combined, and a second maps within individual material classes. This stratification is needed to surface scientifically meaningful hypotheses within single material systems, since cross-material comparisons will be confounded by material identity. The Pipeline requires hypotheses to be supported by at least three samples to be considered further.

In a second step, an LLM reasons over each evidence table, deliberately constrained to consider only the data it contains to prevent introduction of information beyond what the corpus supports. The model cites specific samples, attempts self-falsification, and is provided with any findings already approved by human review as context.

A third step produces a structured finding for each hypothesis, consisting of a summary, a confidence score, identified supporting and complicating records, and questions for review.  A human review provides the final decision --- approve, reject, or defer pending additional data. Approved findings (positive, negative, and inconclusive) are written to an independent append-only ledger and displayed in the Explorer Findings tab.

\section*{Pipeline Demonstration: the Superconducting Qubit Materials Explorer}
The Pipeline demonstrated here visualizes the recent superconducting qubit materials literature produced by the Co-design Center for Quantum Advantage.\cite{c2qa} Among a broad research portfolio, C2QA conducts materials synthesis and characterization and device research across many partner institutions, studying candidate superconducting thin-films for use in quantum computing. This research generates a common set of measured properties (e.g., superconducting transition temperature, electrical resistivity, dielectric loss tangent, resonator internal quality factor, qubit lifetime), but there has been no shared resource to compare results across groups, identify coverage gaps, or test correlations against the published record. Within the superconducting device community, manually curated databases of designs have revealed a community appetite for shared structured data,\cite{squadds2024} but there is no comparable resource for experimental materials characterization.  The corpus described here has been assembled from C2QA publications during the past year.

The C2QA corpus currently comprises 115 digested publications, of which 35 passed relevance triage and contributed 233 sample records across 10 named superconducting material classes (summarized in Table~\ref{tbl:coverage}), 
including tantalum, aluminum, rhenium, niobium, niobium diselenide, and several alloys. The skip rate of ${\sim}65\%$ reflects the breadth of C2QA research, which also spans diamond and neutral atom qubits, quantum algorithms and error correction, and modular computer architectures. The Pipeline identifies and excludes publications outside the scope of the schema without incurring the cost of full ingestion.

The coverage of measured properties varies widely, reflecting a distribution of measurement practices across publications and the Center. For intrinsic material properties, critical temperature is reported for 32\% 
of samples and residual resistance ratio for 12\%. For device-level measurements, 
internal quality factor $Q_i$ is reported for 12\% of samples and qubit 
lifetime $T_1$ for 51\%. These coverage statistics give valuable information on where the collective measurement record is dense versus sparse.

\begin{table}[ht]
\small
\caption{\ Sample counts and measurement coverage by material class in the C2QA corpus. Coverage columns show the number of samples with each property reported. The table covers 197 samples across 10 named material classes. An additional 36 sample records in the corpus either represent other materials or did not report the film material.}
\label{tbl:coverage}
\begin{tabular*}{0.48\textwidth}{@{\extracolsep{\fill}}lrrrrr}
\hline
\textbf{Material} & \textbf{Samples} & \textbf{$T_c$} & 
\textbf{RRR} & \textbf{$Q_i$} & \textbf{$T_1$} \\
\hline
Ta            & 111 & 29 & 9  & 2  & 74 \\
Al            & 26  & 9  & 9  & 0  & 15 \\
Re            & 13  & 1  & 1  & 12 & 6  \\
NbSe$_2$      & 12  & 6  & 0  & 12 & 0  \\
PtSi          & 11  & 11 & 0  & 0  & 0  \\
Ta-Hf         & 8   & 8  & 8  & 1  & 0  \\
Mo$_3$Al$_2$C & 6   & 6  & 0  & 0  & 0  \\
NbN           & 5   & 4  & 0  & 2  & 0  \\
Nb            & 3   & 0  & 0  & 0  & 0  \\
TaN           & 2   & 0  & 0  & 0  & 0  \\
\hline
\end{tabular*}
\end{table}

The Explorer is hosted publicly at \url{https://c2qa-materials-explorer.onrender.com} and is updated as new C2QA publications are digested.

\begin{figure}[h!]
\centering
\includegraphics[width=\columnwidth]{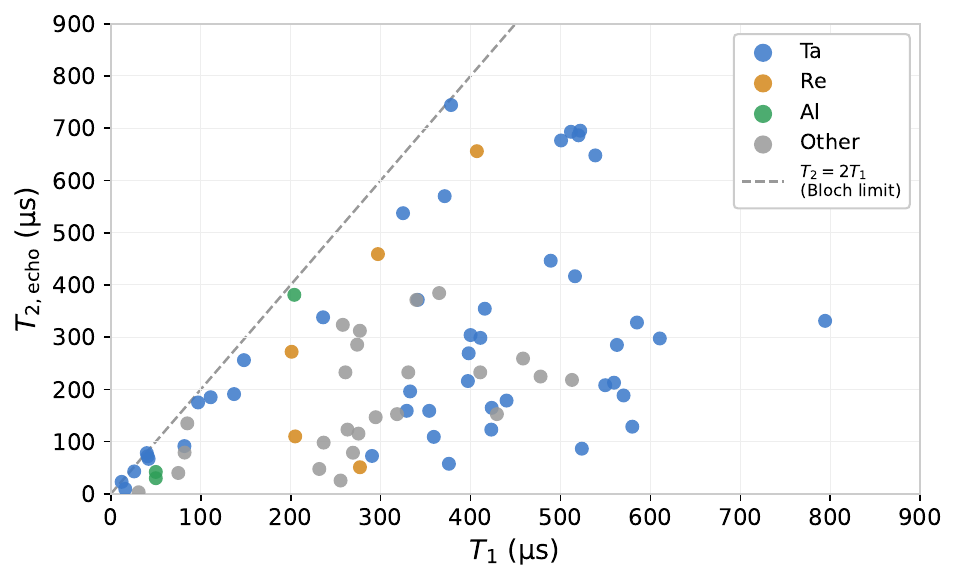}
\caption{$T_{2,\mathrm{echo}}$ vs.\ $T_1$ for all samples 
with both quantities reported, colored by material. The dashed line shows 
the Bloch limit $T_2 = 2T_1$.}
\label{fig:t1t2}
\end{figure}

The Explorer naturally visualizes cross-publication observations that are difficult to realize manually. For example, Fig.~\ref{fig:t1t2} shows $T_{2,\mathrm{echo}}$ versus $T_1$ for 78 qubit samples from 11 C2QA studies reporting both quantities. At smaller $T_1$, $T_{2,\mathrm{echo}}$ tracks near the Bloch limit $T_2 = 2T_1$, consistent with energy relaxation being the primary performance limiter. However, as $T_1$ has improved beyond ${\sim}200$~\textmu s through materials improvements, $T_{2,\mathrm{echo}}$ has not kept pace, falling increasingly below the Bloch limit, indicating that dephasing is limiting qubit performance. 

For this demonstration, corpus mining surfaced 41 author-stated correlations from the catchall blocks of corpus records. Of these, 27 were set aside as out of scope or having insufficient evidence, with 14 proceeding to AI reasoning that produced 19 findings for human review. Three findings are presented here to illustrate the range of possible outputs (Table~\ref{tbl:findings}). More complete reports are provided in the Supplementary Information (S3).

\begin{table*}[!t]
\small
\caption{\ Summary of corpus mining findings from the C2QA superconducting qubit
materials corpus. Three hypotheses are shown, drawn from author-stated correlations
surfaced from catchall blocks across the corpus. Each finding was assessed by the
pipeline and reviewed by a human scientist. Assessment confidence scores reflect
the strength of evidence in the corpus, at current corpus size.}
\label{tbl:findings}
\begin{tabular*}{\textwidth}{@{\extracolsep{\fill}}p{3.0cm} r p{2.8cm} p{6.2cm} p{2.4cm}}
\hline
\textbf{Hypothesis} & \textbf{\#} & \textbf{Materials} & \textbf{Conclusion} & \textbf{Assessment (confidence)} \\
\hline
Mean free path predicts vortex activation $T$ in Ta films
  & 8 & Ta
  & Dirty-limit films show $\sim$10$\times$ higher $T_\mathrm{act}$ than clean-limit; regime-conditional mapping supported
  & Positive (0.70) \\
\hline
Deposition $T$ negatively associated with $T_c$ across superconducting films
  & 28 & NbN, Re, Ta, Ta-Hf (83:17), Al
  & No cross-material trend; $T_c$ set primarily by material identity; weak within-Ta-Hf signal warrants follow-up
  & Inconclusive (0.28) \\
\hline
Film thickness predicts $T_c$ across material systems
  & 43 & NbN, Ta, Ta-Hf (83:17), Al, PtSi, NbSe$_2$, Re
  & No cross-material relationship; NbSe$_2$ 2D effect correctly isolated as material-specific
  & Not supported (0.15) \\
\hline
\end{tabular*}
\end{table*}

The first result is an analysis of a single study and represents a sanity check. Across eight Ta thin-film samples,\cite{bahrami2026} dirty-limit samples show vortex activation temperatures of 1.7--4.2~K, while clean-limit samples cluster at 0.30--0.57~K, a separation of roughly $10\times$. The mining pipeline correctly identified this separation, confirmed it against the supporting sample measurements, and noted that the relationship between mean free path and activation temperature is non-monotonic. This result reproduces a conclusion drawn by the publication authors, and confirms that the pipeline's reasoning functions correctly.

The second result draws on 28 samples across six materials from multiple independent studies. \cite{yang2026,hedrick2026,potluri2025,bland2025,wang2026,bottcher2025} The Pipeline tested whether deposition temperature is correlated with superconducting $T_c$ across the corpus. The conclusion, that $T_c$ is set primarily by material identity but that deposition temperature can account for 0.4~K variation within one material series, is a cross-corpus result that no single publication could have established. The finding was assessed inconclusive for the general hypothesis, with the small identified single-material trend in Ta-Hf alloys recommended for follow-up as the corpus grows.

The third finding draws from 43 samples across seven material systems. \cite{zaman2026,yang2026,hedrick2026,nanayakkara2026,wang2026,wu2026,bottcher2025,potluri2025,bland2025} The Pipeline found that no cross-material relationship between film thickness and $T_c$ is supported. The pipeline correctly identified that the one clean signal, a monotonic $T_c$ increase from 3.5~K to 7.0~K as NbSe$_2$ film thickness increases from 2 to 10~nm, is a known 2D material confinement effect, and declined to generalize to other material systems. No within-material $T_c$ variation with thickness exists in the corpus, with apparent signals complicated by experimental differences including deposition method, substrate, and annealing conditions. This result demonstrates that the mining pipeline has sufficient domain grounding to distinguish a material-specific physical effect from spurious cross-material patterns.

To assess Pipeline accuracy, ten sample records drawn from distinct publications were compared against their source papers by a domain expert. Fuller results of these comparisons are provided in the Supplementary Information (S4). For each record, ten confidence-tagged schema fields were evaluated, along with all catchall items. Across the ten records, 41 of the 100 schema fields contained populated values with assigned confidence levels. The remaining 59 fields were null, indicating the field was not reported in the source paper. All 59 null assignments were confirmed appropriate.

Of the 41 populated values, 40 were confirmed as correct and one was ambiguous. The ambiguous result was correctly flagged as medium confidence by the Pipeline, demonstrating that the confidence framework functions as designed.

The verification also revealed that the Ingester located and correctly assigned values reported outside the main text. Values were drawn from figure captions, supplementary tables, and appendices, locations that required reading the full document rather than relying on prominent narrative statements. In several cases the correct value appeared only once in the paper (e.g., in a table row or caption). Several of the publications used in the verification test included many distinct samples.  In these cases the Ingester correctly segmented the paper into separate sample records, associating measurements with the correct sample in each.

The catchall fields were verified across all ten records, with 108 of 110 items confirmed against the source documents.

\section*{Generalizability}
The Pipeline architecture described here can be applied to any scientific domain in which structured measurements are reported across a body of publications. It can be deployed to digest any body of literature after defining a schema encoding the measurements of interest and pairing it with an ingestion prompt specifying the relevant material types and properties. The Ingester, Explorer, similarity search, and corpus mining stages operate on the record structure rather than any specific scientific content, requiring no domain-specific modification. As a result, adapting the Pipeline to a new field is primarily a schema design task. Only the schema, ingestion prompt, and similarity profile vocabulary need to be adapted.

\section*{Discussion}
The relatively small C2QA superconducting qubit materials corpus shows the substantial potential benefits of a larger, community-maintained corpus. This work has shown end-to-end Pipeline operation at a modest, community scale. The corpus mining results are a starting point, with most hypotheses returning inconclusive findings so far. It is possible that a larger and growing corpus will accumulate more evidence on each hypothesis, so that mining surfaces more conclusive insights.

Notably, the structured corpus produced by the Pipeline is useful for humans and AI agents alike. While existing published measurements and literature are sometimes considered unsuitable for use by AI agents, here we have shown that the Pipeline can digest the distributed scientific record into a structured, queryable database with full provenance, in a format suitable for AI-driven discovery workflows.

Two limitations of the current Pipeline are worth noting. First, data reported solely in paper figures may be missed or recorded with lower confidence, because the Ingester is not specialized for figure parsing. However, because of the modular Pipeline architecture, a future, more capable figure-reading stage could be integrated in a straightforward way. This same modularity applies to future improvements to the Ingester prompt, similarity profile generation, and corpus mining reasoning. A second limitation at present concerns the Explorer's treatment of non-peer-reviewed data, which is a substantial yet-to-be-tapped scientific resource. Because the current Pipeline provenance framework is built around the published record, extending to supplementary sources, published or unpublished, will require a thoughtful approach to inherited credibility.

One practical lesson from applying the Pipeline to the C2QA corpus concerns identifying the appropriate level of schema granularity. In this implementation, raw extracted values for fields such as superconducting and substrate materials, and deposition method, were initially too specific to support meaningful data visualization, with many extracted string variants effectively representing the same category. The solution was to add a normalization layer, which maps extracted information to canonical short lists suitable for grouping and comparison. This step was applied at database build time rather than at extraction, to preserve the original extracted information as a primary artifact (JSON). A similar lesson applies to measurement variants, with quantities such as resonator internal quality factor and qubit coherence time reported in multiple forms across the literature.  Here again, the solution was to group these and select the best-available derived value for corpus-wide visualizations.

For C2QA, the structured superconducting qubit materials corpus supports the Center's co-design objectives of connecting material properties to qubit performance, and providing materials-level inputs that can ultimately feed quantum circuit resource estimates. These goals have been a primary scientific purpose motivating conceptualization and construction of the Pipeline.  

A longer-term vision for the Materials Explorer Pipeline is a network of domain-specific databases, each maintained by the communities that use them. The common PUK record format and Pipeline architecture would allow measurements, observations, and hypotheses reported across many publications to be examined as a single body of evidence.

\section*{Author contributions}
C.T.\ Black: conceptualization, methodology, software development, manuscript writing: original draft, review, and editing.
\section*{Conflicts of interest}
The author has no conflicts to declare.
\section*{Data availability}
The Materials Explorer Pipeline code is available from the corresponding author upon request. The C2QA corpus is publicly accessible at \url{https://c2qa-materials-explorer.onrender.com}. The papers used in this demonstration are sourced from arXiv and the DOE PAGES database.
\section*{Acknowledgements}
This work was supported by the U.S.\ Department of Energy, Office of Science, National Quantum Information Science Research Centers, Co-design Center for Quantum Advantage, under Contract DE-SC0012704. The pipeline software described in this work was developed with the assistance of Claude (Anthropic). Claude was also used to assist in drafting portions of this manuscript. The author thanks K.~G.\ Yager for many enlightening discussions about the opportunities for AI to transform the practice of scientific research.

\bibliographystyle{rsc}
\bibliography{references}

\newpage

\newpage
\setcounter{page}{1}
\renewcommand{\thepage}{S\arabic{page}}

\begin{center}
{\large\textbf{Supplementary Information}}\\[0.3em]
{\normalsize A General Pipeline for Digesting Scientific Literature into a Shared Scientific Knowledge Base}\\[0.2em]
{\normalsize C.T.\ Black, Co-design Center for Quantum Advantage, Brookhaven National Laboratory}
\end{center}
\vspace{1em}

\section*{S1\quad Example Portable Unit of Knowledge (PUK)}
A complete six-block record for sample Wang\_2026\_Transmon\_1,
extracted from Ref.\,\citenum{wang2026} by the Ingester, is shown
below. Fields not reported in the source publication are omitted.
Block~3 measurement values carry confidence levels (high/medium/low)
and source citations reflecting the Ingester's assessment of how
clearly each value is supported in the source publication. The
similarity profile (Block~6) was generated by the Ingester in Pass~3.

\subsection*{Block 1 — Record Metadata}
\begin{tabular}{@{}ll@{}}
\toprule
\textbf{Field} & \textbf{Value} \\
\midrule
Sample ID          & Wang\_2026\_Transmon\_1 \\
Center             & C2QA \\
Source DOI         & 10.48550/arXiv.2603.11188 \\
Extraction method  & Literature ingestion (Pass 2) \\
Human reviewed     & No \\
Human approved     & No \\
\bottomrule
\end{tabular}

\subsection*{Block 2 — Sample Description}
\begin{tabular}{@{}ll@{}}
\toprule
\textbf{Field} & \textbf{Value} \\
\midrule
Film material          & Re \\
Film thickness         & 150~nm \\
Deposition method      & DC magnetron sputtering \\
Deposition temperature & 900$^\circ$C \\
Annealing temperature  & 1200$^\circ$C \\
Substrate             & Sapphire, c-plane, HEM, annealed \\
Junction              & Al/AlOx/Al, 50~nm \\
\bottomrule
\end{tabular}
\vspace{0.3em}
\noindent{\small\textit{Sources: Section~II paragraph~2; Appendix~A paragraphs~1--2.}}

\subsection*{Block 3 — Structured Measurements}
\begin{tabular}{@{}p{4.5cm}lll@{}}
\toprule
\textbf{Field} & \textbf{Value} & \textbf{Units} & \textbf{Conf.} \\
\midrule
$T_c$                         & 1.9   & K   & medium \\
RRR                           & 19    & —   & high \\
$Q_i$ internal quality factor & $9.3 \times 10^6$ & — & high \\
$T_1$                         & 297   & µs  & high \\
$T_{2,\mathrm{echo}}$         & 459   & µs  & high \\
\bottomrule
\end{tabular}
\vspace{0.3em}
\noindent{\small\textit{Sources: $T_c$: Appendix~B paragraph~1. RRR: Figure~4 caption. $Q_i$: Table~I. $T_1$: Table~I. $T_{2,\mathrm{echo}}$: Table~VIII.}}
\vspace{0.3em}
\noindent All other Block~3 fields (loss tangent, gate fidelity, noise characterization, inter-module properties) were not reported in this publication.

\subsection*{Block 4 — Derived Quantities}
\begin{tabular}{@{}ll@{}}
\toprule
\textbf{Field} & \textbf{Value} \\
\midrule
Derived material  & Re      \\
Derived substrate & Sapphire \\
Derived $Q_i$     & $9.3 \times 10^6$ \\
Derived $T_2$     & 459~µs  \\
$T_1$ context     & qubit\_state \\
\bottomrule
\end{tabular}

\subsection*{Block 5 — Catchall}
Two items were captured in the catchall block:

\textit{Additional measurement:} Ramsey decoherence time $T_{2,\mathrm{Ramsey}} = 42 \pm 19$~µs (Table~VIII). $T_{2,\mathrm{Ramsey}}$ much shorter than $T_{2,\mathrm{echo}}$ suggests significant low-frequency dephasing noise, likely from TLS fluctuators in dielectric interfaces; the ratio $T_{2,\mathrm{echo}}/T_{2,\mathrm{Ramsey}}$ indicates dephasing is not purely from energy relaxation.

\textit{Free notes:} Transmon~1 operates at 5.00~GHz. Junction inductance ${\sim}9$~nH from Dolan bridge double-angle evaporation. Substrate treated with piranha solution before Re deposition; wafer treated with concentrated H$_2$SO$_4$ at 100$^\circ$C and 10:1 BOE after Re patterning before junction fabrication.

\subsection*{Block 6 — Similarity Profile}
\begin{tabular}{@{}ll@{}}
\toprule
\textbf{Dimension} & \textbf{Value} \\
\midrule
Material class   & Rhenium \\
Transport regime & Intermediate \\
Loss mechanisms  & TLS\_interface, TLS\_metal\_vacuum, \\
                 & dielectric\_substrate \\
Device type      & Transmon \\
Coherence tier   & State of the art \\
Science focus    & Device demo, loss identification \\
Growth method    & Sputtering \\
Key correlations & Surface oxide to TLS; $Q_i$ to $T_1$ \\
Profile version  & 1.0 \\
\bottomrule
\end{tabular}

\section*{S2\quad Block~3 Schema Fields — Structured Measurements}
Block~3 of the schema contains defined measurement fields with known
or approximate connections to device performance. Fields marked
with $\dagger$ were promoted from the catchall block based on
measurement frequency across the corpus (May 2026).

\bigskip
\noindent\textbf{Block~3.1 — Superconducting properties:}
$T_c$; $T_c$ uniformity; RRR; sheet resistance; sheet kinetic
inductance$^\dagger$; mean free path$^\dagger$; vortex activation
temperature$^\dagger$; London penetration depth; upper critical field.

\bigskip
\noindent\textbf{Block~3.2 — Dielectric and surface loss:}
Loss tangent (substrate); loss tangent (interface); TLS density;
TLS coupling strength; surface oxide thickness; surface oxide
composition; surface participation ratio.

\bigskip
\noindent\textbf{Block~3.3 — Microwave performance:}
$Q_i$ internal; $Q_i$ single photon; $Q_c$ coupling;
$Q_{\mathrm{TLS},0}$ (unsaturated); resonator type; resonator
gap width; $p_{MS,\mathrm{resonator}}$; $p_{MS,\mathrm{pad}}$;
loss mechanism.

\bigskip
\noindent\textbf{Block~3.4 — Qubit performance:}
$T_1$; $T_{2,\mathrm{echo}}$; $T_{2,\mathrm{Ramsey}}$; qubit
frequency; anharmonicity; single-qubit gate fidelity; single-qubit
gate time; two-qubit gate fidelity; two-qubit gate time; readout
fidelity; readout time.

\bigskip
\noindent\textbf{Block~3.5 — Noise characterization:}
Flux noise; charge noise; $1/f$ noise exponent; quasiparticle
density; quasiparticle parity switching rate.

\section*{S3\quad Corpus Mining Findings — Full Reports}
Three corpus mining findings are summarized in the main text. Full reports for each are provided below, in the format produced by the Pipeline's Phase~C write-up stage. Each finding includes the hypothesis tested, a summary assessment, supporting and complicating records, detailed analysis, alternative explanations, and recommended next steps.

Finding types: $\checkmark$~positive \quad $\times$~not supported \quad $\sim$~inconclusive

\subsection*{Finding S1 \quad $\checkmark$~positive \quad Confidence: 70\%}
\textbf{Dirty-limit Ta thin films exhibit vortex activation temperatures roughly one order of magnitude higher than clean-limit films from the same study.}

\noindent\textit{Samples:} 8 \quad \textit{Materials:} Ta

\subsubsection*{Summary}
Across 8 Ta thin-film samples from a single study,\cite{bahrami2026}, dirty-limit samples (mean free path $l = 12.9$--$20.4$~nm) show vortex activation temperatures of 1.69--4.17~K, while all five clean-limit samples ($l = 81.5$--$142.3$~nm) cluster at 0.29--0.57~K --- a separation of roughly 3--14$\times$ in $T_\mathrm{act}$. The between-regime signal is large and internally replicated, but within the clean-limit cluster the correlation between $l$ and $T_\mathrm{act}$ is weak and non-monotonic, indicating that mean free path alone does not continuously predict vortex activation temperature at fine resolution. All data originate from a single material (Ta), single research group, and single deposition method, which substantially limits generalizability.

\subsubsection*{Supporting records}
\begin{itemize}[leftmargin=*, itemsep=1pt, topsep=2pt]
  \item Bahrami\_2026\_D7 ($l = 12.9$~nm, $T_\mathrm{act} = 4.16$~K)
  \item Bahrami\_2026\_D8 ($l = 13.3$~nm, $T_\mathrm{act} = 4.17$~K)
  \item Bahrami\_2026\_D6 ($l = 20.4$~nm, $T_\mathrm{act} = 1.69$~K)
  \item Bahrami\_2026\_D2, Bahrami\_2026\_D3
\end{itemize}

\subsubsection*{Complicating records}
\begin{itemize}[leftmargin=*, itemsep=1pt, topsep=2pt]
  \item Bahrami\_2026\_D3, Bahrami\_2026\_D1, Bahrami\_2026\_D5
\end{itemize}

\subsubsection*{Detail}
The dataset separates cleanly into two regimes at approximately $l \sim 40$--$80$~nm. Dirty-limit samples D7 ($l = 12.9$~nm, $T_\mathrm{act} = 4.16$~K), D8 ($l = 13.3$~nm, $T_\mathrm{act} = 4.17$~K), and D6 ($l = 20.4$~nm, $T_\mathrm{act} = 1.69$~K) all exceed the highest clean-limit $T_\mathrm{act}$ (D5: 0.57~K) by factors of 3--7$\times$, and the two dirtiest samples exceed it by ${\sim}7\times$. The clean-limit samples D1--D5 ($l = 81.5$--$142.3$~nm) span only 0.29--0.57~K in $T_\mathrm{act}$ despite a ${\sim}1.7\times$ range in $l$, and the ordering is not monotonic (e.g., D4 at $l = 125.2$~nm has $T_\mathrm{act} = 0.33$~K while D5 at $l = 112.2$~nm has $T_\mathrm{act} = 0.57$~K).

Both mean free path and vortex activation temperature are model-derived quantities: $l$ is typically computed from resistivity and RRR using bulk scattering parameters that may not apply to thin films, and $T_\mathrm{act}$ is extracted from thermally-activated flux-flow fits. The magnitude of the between-regime separation (${\sim}$one order of magnitude) is large enough that moderate systematic errors in either derived quantity are unlikely to eliminate the signal, but this model-dependency is a non-trivial caveat. The single-study, single-material, single-deposition-method provenance of all 8 samples means the pattern could reflect a specific process parameter (e.g., sputtering pressure) that co-varies with both $l$ and $T_\mathrm{act}$ without a direct causal link.

\textit{Supporting:} The two dirtiest samples (D7: $l = 12.9$~nm, $T_\mathrm{act} = 4.16$~K; D8: $l = 13.3$~nm, $T_\mathrm{act} = 4.17$~K) are near-identical replicates at the dirty-limit extreme, providing internal validation that the elevated $T_\mathrm{act}$ is not a single-sample artifact. The intermediate dirty-limit sample (D6: $l = 20.4$~nm, $T_\mathrm{act} = 1.69$~K) sits between the two regimes in a manner consistent with a monotonic trend across the dirty/clean-limit boundary.

\textit{Complicating:} Within the clean-limit cluster, the $l$--$T_\mathrm{act}$ relationship is non-monotonic: D3 ($l = 81.5$~nm) has the lowest $T_\mathrm{act}$ (0.29~K) despite the shortest clean-limit mean free path, while D5 ($l = 112.2$~nm) has the highest clean-limit $T_\mathrm{act}$ (0.57~K) despite not having the longest $l$. This scatter indicates that additional variables modulate $T_\mathrm{act}$ within the clean limit independently of mean free path.

\subsubsection*{Alternative explanation}
The most plausible alternative is that the true driver is microstructural disorder (defect and grain boundary density) rather than mean free path per se: disorder simultaneously reduces $l$ and increases pinning center density, raising vortex activation energy. In this interpretation, $l$ is a proxy for disorder rather than the direct physical cause of $T_\mathrm{act}$ differences, and the regime-level separation reflects a qualitative change in vortex pinning physics (e.g., collective vs.\ single-vortex creep) rather than a smooth continuous dependence on $l$.

\subsubsection*{Recommended next steps}
Ta is the only material in this corpus with data for this hypothesis, with 8 samples sufficient to warrant a focused re-analysis for Ta specifically --- particularly to examine whether the regime-level separation holds across different deposition methods and substrates. Re-analysis for Nb, NbN, and Al would be valuable if samples with both mean free path and vortex activation temperature measurements can be identified, but cannot be recommended on current data.

\subsection*{Finding S2 \quad $\sim$~inconclusive \quad Confidence: 28\%}
\textbf{Deposition temperature shows a weak, material-specific negative association with $T_c$ in Ta-Hf alloy films but no consistent cross-material trend across six superconducting thin-film systems.}

\noindent\textit{Samples:} 28 \quad \textit{Materials:} NbN, Re, Ta, Ta-Hf (83:17), Al

\subsubsection*{Summary}
Across 28 samples spanning NbN, Re, Ta, Ta-Hf (83:17), and Al, no consistent cross-material relationship between deposition temperature and $T_c$ is observed; material identity dominates $T_c$ by ${\sim}10$~K, far exceeding any processing-temperature effect. The only controlled within-material series (Ta-Hf 83:17, Yang\_2026, $n = 8$ across four temperatures from 550--850$^\circ$C) shows a weak negative trend of ${\sim}0.37$~K over 300$^\circ$C, but this trend is non-monotonic at intermediate temperatures and physically modest. The general cross-material hypothesis is not supported; the Ta-Hf-specific observation warrants cautious follow-up but cannot be generalized.

\subsubsection*{Supporting records}
\begin{itemize}[leftmargin=*, itemsep=1pt, topsep=2pt]
  \item Yang\_2026\_Table\_2\_Row\_1 (Ta-Hf, 550$^\circ$C, $T_c = 6.17$~K)
  \item Yang\_2026\_Table\_2\_Row\_5 (Ta-Hf, 550$^\circ$C, $T_c = 6.21$~K)
  \item Yang\_2026\_Table\_2\_Row\_4 (Ta-Hf, 850$^\circ$C, $T_c = 5.84$~K)
  \item Yang\_2026\_Table\_2\_Row\_8 (Ta-Hf, 850$^\circ$C, $T_c = 6.0$~K)
  \item Hedrick\_2026\_Table\_II\_Row\_3 (Al, $-72^\circ$C, $T_c = 1.15$~K)
\end{itemize}

\subsubsection*{Complicating records}
\begin{itemize}[leftmargin=*, itemsep=1pt, topsep=2pt]
  \item Hedrick\_2026\_Table\_II\_Row\_8 and Row\_9 (Al, Si substrate, 25$^\circ$C, $T_c = 1.41$~K)
  \item Potluri\_2025\_Table\_I\_Row\_1 (Ta, Si, 22$^\circ$C, $T_c = 4.4$~K) vs.\ Row\_2 (Ta, Al$_2$O$_3$, 500$^\circ$C, $T_c = 4.3$~K)
  \item Bland\_2025\_Ta-on-Si (Ta, Si, 625$^\circ$C, $T_c = 4.2$~K)
  \item Wang\_2026\_Re\_film and Transmon samples (Re, 900$^\circ$C, $T_c = 1.9$~K)
  \item Yang\_2026\_Table\_2\_Row\_2 (Ta-Hf, 650$^\circ$C, $T_c = 6.2$~K)
  \item Yang\_2026\_Table\_2\_Row\_6 (Ta-Hf, 650$^\circ$C, $T_c = 6.18$~K)
\end{itemize}

\subsubsection*{Detail}
The 28-sample corpus spans six material families deposited by at least three distinct methods (ALD, e-beam evaporation, DC sputtering), making cross-material regression on deposition temperature statistically and physically unjustified. Within Ta-Hf (83:17) on sapphire (Yang\_2026), $T_c$ declines from 6.21~K at 550$^\circ$C to 5.84~K at 850$^\circ$C, but the 650$^\circ$C replicates ($T_c = 6.18$--$6.20$~K) slightly exceed some 550$^\circ$C values, indicating the trend is non-monotonic and the ${\sim}0.37$~K total variation is comparable to sample-to-sample scatter.

For Ta (Potluri\_2025 and Bland\_2025), $T_c$ spans only 4.2--4.4~K across deposition temperatures of 22$^\circ$C to 625$^\circ$C on different substrates (Si and Al$_2$O$_3$), with no discernible trend and substrate identity as an uncontrolled confound. Re films (Wang\_2026) are all deposited at 900$^\circ$C ($T_c = 1.9$~K), providing no within-material variation, and post-deposition annealing at 1200$^\circ$C for Transmon samples likely dominates microstructure. NbN (B\"{o}ttcher\_2025) contributes only a single deposition temperature (400$^\circ$C, $T_c = 12.0$~K via ALD), offering no within-material leverage. The Al dataset (Hedrick\_2026) conflates deposition temperature with substrate identity and transport regime, rendering it uninterpretable for this hypothesis without additional controls.

\textit{Supporting:} Within the Ta-Hf (83:17) series (Yang\_2026), $T_c$ is highest at 550$^\circ$C (6.17--6.21~K) and lowest at 850$^\circ$C (5.84--6.0~K), consistent with a weak negative deposition temperature--$T_c$ relationship within this single alloy system on sapphire substrates.

\textit{Complicating:} Cross-material comparison is entirely dominated by intrinsic material properties: NbN deposited at 400$^\circ$C yields $T_c = 12$~K while Re deposited at 900$^\circ$C yields $T_c = 1.9$~K, making any cross-material regression on deposition temperature physically meaningless. Within Al (Hedrick\_2026), apparent $T_c$ differences between deposition temperatures are confounded by substrate changes (c-Al$_2$O$_3$ vs.\ Si) and transport regime, preventing attribution to deposition temperature alone.

\subsubsection*{Alternative explanation}
The most plausible alternative is that deposition temperature acts indirectly on $T_c$ through microstructural intermediaries --- crystalline phase selection, grain size, residual stress, or alloy homogeneity --- rather than through a direct thermodynamic effect on the superconducting condensate. This would explain why the effect is material-specific, modest in magnitude, and non-monotonic. Substrate identity, film thickness, deposition rate, and post-deposition annealing are all plausible dominant confounds that co-vary with deposition temperature across this corpus.

\subsubsection*{Recommended next steps}
This hypothesis should be re-run per material class rather than cross-material. Ta-Hf (83:17) is the only material with sufficient within-material variation ($n = 8$, four temperature points, two replicates each) to warrant a focused re-analysis and is recommended for immediate follow-up. Ta has three samples across a wide temperature range (22--625$^\circ$C) but on different substrates, making it a candidate for re-analysis only if substrate is treated as a covariate or if additional same-substrate samples can be identified. Al, Re, and NbN each have either single deposition temperatures or substrate confounds that make per-material re-analysis premature without new data.

\subsection*{Finding S3 \quad $\times$~not supported \quad Confidence: 15\%}
\textbf{No cross-material correlation between film thickness and superconducting critical temperature is supported; a within-material effect is confirmed only for ultra-thin NbSe$_2$.}

\noindent\textit{Samples:} 43 \quad \textit{Materials:} NbN, Ta, Ta-Hf (83:17), Al, PtSi, NbSe$_2$, Re

\subsubsection*{Summary}
Across 43 samples spanning nine material systems, no coherent cross-material relationship between film thickness and $T_c$ is observed; material identity, deposition method, substrate, and post-deposition annealing dominate $T_c$ variation. The only well-supported within-material thickness--$T_c$ relationship is in NbSe$_2$ (6 samples, 2--10~nm), where $T_c$ rises monotonically from 3.5~K to 7.0~K as thickness increases --- a known 2D superconductor quantum-confinement effect, not a generic thin-film phenomenon. For all other materials in this corpus (NbN, Ta, Ta-Hf, Al, PtSi, Re), either no thickness variation exists within the material or the apparent thickness signal is confounded by process and substrate differences.

\subsubsection*{Supporting records}
\begin{itemize}[leftmargin=*, itemsep=1pt, topsep=2pt]
  \item Zaman\_2026\_D1 through D5, D8 (NbSe$_2$, 2--10~nm thickness series)
  \item Yang\_2026\_Table\_2\_Row\_1 through Row\_4 (Ta-Hf, 250~nm) vs.\ Row\_5 through Row\_8 (Ta-Hf, 470~nm)
\end{itemize}

\subsubsection*{Complicating records}
\begin{itemize}[leftmargin=*, itemsep=1pt, topsep=2pt]
  \item Hedrick\_2026\_Table\_II (Al, all at 200~nm --- no thickness variation)
  \item Nanayakkara\_2026 (PtSi, all at 25~nm)
  \item Wang\_2026\_Re\_film\_characterization\_sample (Re, all at 150~nm)
  \item Wu\_2026\_M2M\_converter (NbN, 50~nm, 1000$^\circ$C anneal)
  \item B\"{o}ttcher\_2025\_NbN\_CPW\_resonators (NbN, ALE-thinned)
  \item Potluri\_2025\_Table\_I\_Row\_1 (Ta, 100~nm)
  \item Yang\_2026\_Table\_2\_Row\_5 through Row\_8 (Ta-Hf, 470~nm)
\end{itemize}

\subsubsection*{Detail}
The NbSe$_2$ result (Zaman\_2026 D1--D8) is the only clean signal in the dataset: six samples spanning 2--10~nm show a monotonic $T_c$ increase (3.5 $\to$ 7.0~K) consistent with suppression of superconductivity in the 2D ultra-thin limit, a well-understood phenomenon driven by enhanced quantum fluctuations and disorder at reduced dimensionality. This effect saturates near 10~nm, consistent with approach to the bulk NbSe$_2$ $T_c$ (${\sim}7.2$~K), and is not generalizable to other materials.

For NbN, the apparent thickness range (5--50~nm across B\"{o}ttcher\_2025 and Wu\_2026) is confounded by different ALD chamber conditions, ALE thinning steps, and a 1000$^\circ$C anneal in one study, making any thickness--$T_c$ inference unreliable. Ta-Hf (83:17) samples at 250~nm and 470~nm (Yang\_2026) show statistically indistinguishable $T_c$ values (mean ${\sim}6.06$~K vs.\ ${\sim}6.11$~K), directly contradicting a meaningful thickness effect in this alloy at these scales. Ta samples across studies (Potluri\_2025 at 100--150~nm, Bland\_2025 at 200~nm) show a marginal apparent $T_c$ decrease (4.4 $\to$ 4.2~K over 100~nm) within the range attributable to substrate differences (Si vs.\ Al$_2$O$_3$) and inter-study variability.

\textit{Supporting:} NbSe$_2$ samples (Zaman\_2026, D1--D5, D8) show a clear, monotonic positive $T_c$--thickness relationship from 3.5~K at ${\sim}2$~nm to 7.0~K at ${\sim}10$~nm, with apparent saturation near bulk $T_c$, consistent with well-established 2D superconductor physics. This is the only material in the corpus with sufficient within-material thickness variation to assess the hypothesis directly.

\textit{Complicating:} All other materials lack within-material thickness variation: all 9 Al samples are at 200~nm, all 6 PtSi samples at 25~nm, and all Re samples at 150~nm, structurally preventing thickness--$T_c$ analysis for the majority of the corpus. NbN shows non-monotonic apparent thickness dependence fully explained by differing deposition methods and annealing conditions rather than thickness per se.

\subsubsection*{Alternative explanation}
Material identity is the dominant determinant of $T_c$ across this dataset, with intrinsic $T_c$ spanning 0.9--12~K reflecting bulk material properties rather than thickness effects. Within individual materials, deposition technique, substrate choice, and post-deposition annealing temperature appear to be far stronger predictors of $T_c$ than film thickness in the 20--200~nm regime where most samples reside --- well above the ultra-thin limits where thickness-driven $T_c$ suppression is physically expected.

\subsubsection*{Recommended next steps}
This hypothesis should be re-run per material class if and when appropriate within-material thickness series data become available. Current sample counts per material with thickness variation: NbSe$_2$ (6 samples, 2--10~nm range --- sufficient, analysis complete); NbN (3 samples across 5--50~nm but with uncontrolled process confounders --- insufficient without process metadata); Ta (3 samples across 100--200~nm but across different studies and substrates --- insufficient); Ta-Hf (8 samples at only two thickness values --- marginal, null result at 250 vs.\ 470~nm is informative); Al, PtSi, Re (no thickness variation --- cannot be re-run). Priority for re-analysis: NbSe$_2$ (data exists, analysis complete), NbN (high scientific interest, needs process-controlled series), Ta (high quantum computing relevance, needs systematic thickness study).

\section*{S3\quad Extraction Accuracy Verification}

Ten sample records were selected for verification against their source
publications. Papers were chosen to span a range of publication types
and to include several with large numbers of characterized samples, to
test whether the Ingester correctly segments multi-sample papers and
associates measurements with the correct sample. The ten samples
verified were: Hedrick\_2026\_Cryo\,\citenum{hedrick2026} (Al CPW
resonator, cryogenic deposition study),
Zaman\_2025\_D3\,\citenum{zaman2026} (NbSe$_2$ kinetic inductance
resonator), Bahrami\_2026\_D3\,\citenum{bahrami2026} (Ta thin film,
transport study), Dai\_2026\_3D\_transmon\,\citenum{zdpg-mhpc} (3D
transmon), Nho\_2026\_Big-delta\,\citenum{ql6q-wfpn} (Al 3D transmon,
quasiparticle gap engineering),
Chang\_2025\_Ta\_Au\,\citenum{PhysRevLett.134.097001} (Ta with Au
encapsulation), Yang\_2026\_Ta-Hf\,\citenum{yang2026} (Ta-Hf alloy,
8-sample deposition series),
Joshi\_2026\_qubit-3\,\citenum{joshi2026}
(Ta transmon, participation ratio study),
Molinelli\_2026\,\citenum{molinelli2026}
(8-qubit Ta processor), and
Bland\_2025\_Qubit\_21\,\citenum{bland2025} (Ta transmon, 57-qubit
processor study).

For each record, the ten confidence-tagged schema fields were
evaluated: $T_c$, RRR, $Q_i$, $T_1$, loss tangent, $T_{2,\mathrm{echo}}$,
surface oxide thickness, film thickness, junction present, and
resonator gap width. The verification protocol used a two-step
procedure. First, for each populated field, a language model was asked
to locate the specific sentence, table entry, or figure caption in the
source paper from which the value was drawn, and to reproduce that
text. Second, the author confirmed that the reproduced text does in
fact support the extracted value. A field was judged \textit{supported}
if the source text unambiguously substantiates the extracted value,
\textit{ambiguous} if the source text is consistent with the value but
does not uniquely determine it, and \textit{not found} if no supporting
text could be located. Null fields (not extracted by the Ingester) were
confirmed by checking that the quantity is indeed absent from the paper.
All catchall items were verified by the same procedure.

Table~\ref{tbl:verification} summarizes results across all ten samples.
Of 100 field assessments (10 records $\times$ 10 fields), 41 fields
were populated and 59 were null. All 40 supported values and all 59
null assignments were confirmed correct. The single ambiguous result
(Molinelli\_2026, $T_1$, medium confidence) arose because the extracted
value appears to be a mean computed across 48 table entries rather than
a value stated explicitly in the paper; the Ingester correctly flagged
this field as medium confidence. No not-found results were recorded
across any of the ten records.

Of 110 catchall items evaluated across the ten records, 108 were
confirmed against their source documents. The two unconfirmed items
(Yang\_2026\_Ta-Hf) could not be located in the version of the paper
available for review and may reflect extraction from a figure or
supplementary section not reproduced in the accessible copy.

\begin{table}[h]
\small
\caption{\ Extraction accuracy verification results. Ten sample records
verified against source publications. Supported: extracted value
confirmed against source text. Ambiguous: source text consistent but
not uniquely determinative. Null (correct): field absent from PUK and
confirmed not reported in source paper.}
\label{tbl:verification}
\resizebox{\columnwidth}{!}{%
\begin{tabular}{@{}lrrrrrr@{}}
\hline
\textbf{Sample} & \textbf{Pop.} & \textbf{Supp.} & \textbf{Amb.} &
\textbf{NF} & \textbf{Null} & \textbf{Catchall} \\
\hline
Hedrick\_2026\_Cryo      & 6 & 6 & 0 & 0 & 4  & 3/3   \\
Zaman\_2025\_D3          & 3 & 3 & 0 & 0 & 7  & 8/8   \\
Bahrami\_2026\_D3        & 4 & 4 & 0 & 0 & 6  & 10/11 \\
Dai\_2026\_3D\_transmon  & 2 & 2 & 0 & 0 & 8  & 27/27 \\
Nho\_2026\_Big-delta     & 3 & 3 & 0 & 0 & 7  & 16/16 \\
Chang\_2025\_Ta\_Au      & 3 & 3 & 0 & 0 & 7  & 13/13 \\
Yang\_2026\_Ta-Hf        & 8 & 8 & 0 & 0 & 2  & 2/3   \\
Joshi\_2026\_qubit-3     & 5 & 5 & 0 & 0 & 5  & 2/2   \\
Molinelli\_2026          & 3 & 2 & 1 & 0 & 7  & 22/22 \\
Bland\_2025\_Qubit\_21   & 4 & 4 & 0 & 0 & 6  & 5/5   \\
\hline
\textbf{Total}           & \textbf{41} & \textbf{40} & \textbf{1} &
\textbf{0} & \textbf{59} & \textbf{108/110} \\
\hline
\end{tabular}}
\end{table}

\noindent Column headers: Pop.\ = fields populated; Supp.\ =
supported; Amb.\ = ambiguous; NF = not found; Null = null assignments
confirmed correct; Catchall = catchall items confirmed / evaluated.

\vspace{1em}
\noindent\rule{\columnwidth}{0.4pt}
\vspace{0.5em}
\noindent{\small \textit{Supplementary Information generated from Materials Explorer Pipeline corpus mining output. C2QA corpus mining run: May 2026. Full corpus accessible at \url{https://c2qa-materials-explorer.onrender.com}}}

\end{document}